\documentclass[%
superscriptaddress, %
twocolumn, %
showpacs,
 amsmath,amssymb,
 aps,
prb,
floatfix, longbibliography ]{revtex4-1}
\usepackage{graphicx}
\usepackage{dcolumn}
\usepackage{bm}
\usepackage{soul,color}

\usepackage[colorlinks,linkcolor=blue,anchorcolor=blue,citecolor=blue,urlcolor=blue]{hyperref}
\usepackage{tabularx}
\usepackage{mathtools}

\newcommand{\mb}[1]{\mathbf{#1}}

\begin{document}

\title{Symmetry breaking in the double moir\'{e} superlattices of relaxed twisted bilayer graphene on hexagonal boron nitride}

\author{Xianqing Lin}
\email[E-mail: ]{xqlin@zjut.edu.cn}
\affiliation{College of Science,
             Zhejiang University of Technology,
             Hangzhou 310023, People's Republic of China}

\author{Jun Ni}
\affiliation{State Key Laboratory of Low-Dimensional Quantum Physics and Frontier Science Center for Quantum Information,
             Department of Physics, Tsinghua University, Beijing 100084,
             People's Republic of China}

\date{\today}

\begin{abstract}
We study the atomic and electronic structures of the commensurate double moir\'{e} superlattices in
fully relaxed twisted bilayer graphene (TBG) nearly aligned with the hexagonal boron nitride (BN).
The single-particle effective Hamiltonian ($\hat{H}^0$) taking into account the relaxation effect and the full
moir\'{e} Hamiltonian introduced by BN has been built for TBG/BN. The mean-field (MF) band structures of the self-consistent Hartree-Fock (SCHF) ground states
at different number ($\nu$) of filled flat bands relative to the charge neutrality point (CNP) are obtained
based on $\hat{H}^0$ in the plane-wave-like basis.
The single-particle flat bands in TBG/BN become separated by the opened gap at CNP due to the symmetry breaking in $\hat{H}^0$.
We find that the broken
$C_2$ symmetry in $\hat{H}^0$ mainly originates from the intralayer inversion-asymmetric
structural deformation in the graphene layer adjacent to BN,
which introduces spatially non-uniform modifications of the intralayer Hamiltonian.
The gapped flat bands have finite Chern numbers.
For TBG/BN with the magic twist angle, the SCHF ground states with $|\nu|$ = 1-3 are all insulating with narrow MF gaps.
When the flat conduction bands are filled, the gap at $\nu$ = 1 is smaller than that at $\nu$ = 3, suggesting that
the nontrivial topological properties associated with the flat Chern bands are more likely to be observed at $\nu = 3$.
This is similar for negative $\nu$ with empty valence bands.
The dependence of the electronic structure of TBG/BN on positive $\nu$ is roughly consistent
with recent experimental observations.
\end{abstract}

\pacs{%
}



\maketitle


\section{Introduction}

The recently realized magic-angle twisted bilayer graphene (TBG) has inspired
great interest in exploring its peculiar
electronic structure associated with the flat bands around the Fermi level
\cite{Cao2018,cao2018unconventional,EmergentSharpe605,lu2019superconductors,uri2020mapping}.
The flat bands in TBG with the magic angle ($\theta_m$) of about 1.1$^\circ$ have a total
capacity of eight electrons
per moir\'{e} supercell\cite{Bistritzer12233,Trambly2012,LopesdosSantos2012,Fang2016,DT268,OriginPhysRevLett.122.106405}, and
the number ($\nu$) of filled flat bands relative to the charge neutrality point (CNP) can be tuned in
the range of $-4 \sim 4$ in experiments\cite{Cao2018,cao2018unconventional,EmergentSharpe605,lu2019superconductors,uri2020mapping}.
The positive and negative $\nu$ are used to represent the number of filled conduction bands and empty
valence bands relative to CNP, respectively.
Correlated insulating states at integer $\nu$ with $|\nu| \leq 3$ were observed in TBG
with $\theta_m$\cite{Cao2018,cao2018unconventional,EmergentSharpe605,lu2019superconductors,uri2020mapping}.
In particular, the quantized anomalous Hall (QAH) effect was realized in TBG aligned with the hexagonal boron nitride
(TBG/BN)\cite{Intrinsic2020Serlin}. The QAH state emerges at $\nu = 3$ and the insulating state at $\nu = 2$ was also observed\cite{Intrinsic2020Serlin}.
Theoretical studies attributed the nontrivial topological properties of TBG/BN to the broken $C_2$ symmetry
induced by BN\cite{Mechanism2020nick,Twisted209zhang,Spontaneous2019liu,zhang2020correlated}.
We note that the theoretical models in these studies only included the uniform sublattice-asymmetric potential in the
graphene layer on BN, while the nonuniform moir\'{e}
potentials introduced by BN have been ignored\cite{Mechanism2020nick,Twisted209zhang,Spontaneous2019liu,zhang2020correlated}.
Moreover, only the rigid TBG or some relaxation effect
in TBG with one empirical parameter were considered\cite{Mechanism2020nick,Twisted209zhang,Spontaneous2019liu,zhang2020correlated}.
However, the rigid double moir\'{e} superlattices in TBG/BN
undergo spontaneous in-plane relaxation and out-of-plane corrugation due to the energy gain from the larger
domains of energetically favorable stacking configurations of graphene bilayers and also of graphene on BN,
similar to the pristine TBG\cite{McEuenBLG13,Uchida2014,Wijk2015,Dai2016,Jain2017,Nam2017,
Gargiulo2018,carr2018relaxation,ShearPhysRevB.98.195432,
Atomicyoo2019atomic,CrucialPhysRevB.99.195419,ContinuumPhysRevB.99.205134} and
the graphene monolayer on BN\cite{Effect2015Oct,Origin2015Feb,Moire2017Aug,Effective2019lin}.
The strong structural deformation in the graphene layer on BN can play an important role in the symmetry breaking
of the single-particle Hamiltonian of the relaxed TBG/BN.
The opened gap at CNP in the graphene monolayer aligned with BN has been found to be greatly enhanced by
the structural relaxation\cite{Effect2015Oct,Origin2015Feb,Moire2017Aug,Effective2019lin}.
Therefore, it is important to take into account the full relaxation of TBG/BN and the full effective Hamiltonian
introduced by BN to describe the electronic structure of TBG/BN associated with the broken $C_2$ symmetry.

The self-consistent Hartree-Fock (SCHF) method can be used to treat the electron-electron interaction in TBG, while
this method was based on the continuum Hamiltonian of
TBG with a rigid superlattice or with only the interlayer-spacing corrugation\cite{Nature2020xie,zhang2020correlated}.
The approach to employing the SCHF method to describe the electronic structure of the fully relaxed TBG/BN remains to be developed.

\begin{figure*}[t]
\includegraphics[width=1.7\columnwidth]{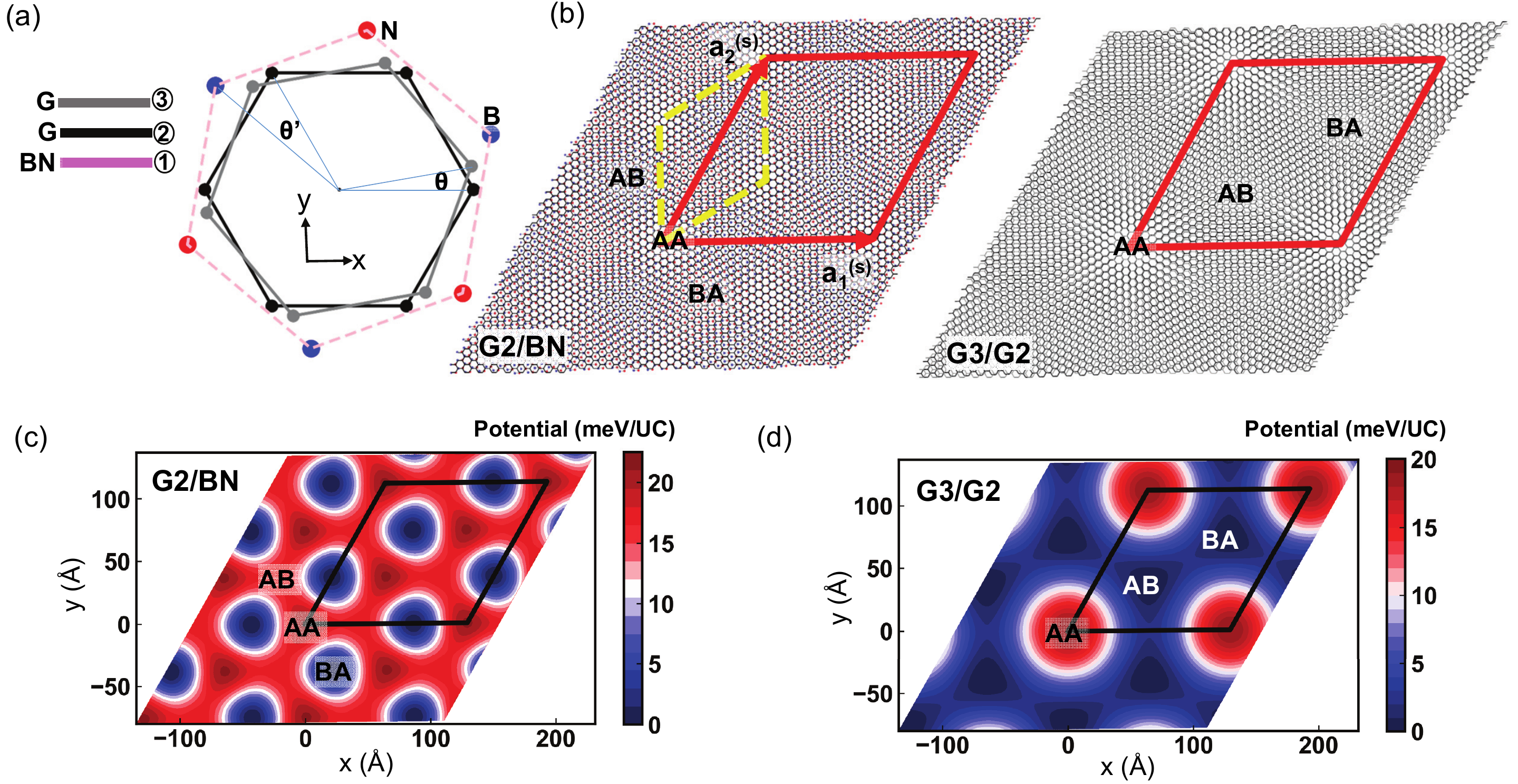}
\caption{(Color online) The structure and interlayer interaction potentials of the rigid double moir\'{e} superlattices in
TBG/BN.
(a) The schematic view of the TBG/BN trilayer. The top graphene layer (G3) and the bottom BN layer are rotated by
$\theta$ and $\theta'$ counterclockwise with respect to the fixed middle graphene layer (G2), respectively.
The B and N atoms are represented by blue and red circles, respectively.
The lattice-constant mismatch between graphene and BN is amplified in this view and its value is about -1.70\%.
(b) The schematic view of the commensurate moir\'{e} superlattices in G2 on BN (G2/BN) and in G3 on G2 (G3/G2).
$\mb{a_j^{(s)}}$ ($j=1,2$) are the basis vectors of the superlattice in G3/G2 and this superlattice is strictly periodic.
An approximate moir\'{e} supercell in G2/BN is indicated by the yellow dashed lines.
In parts of G2/BN with BA (AB) stacking, the sublattice-B (sublattice-A) carbon atoms are above the
boron (nitrogen) atoms, and the carbon atoms in the other sublattice are above the
hexagon centers of BN.
(c) and (d) The spatial distribution of the local interlayer interaction potential as a function of the
position in the rigid superlattices of G2/BN (c) and G3/G2 (d) for commensurate TBG/BN with $\theta = 1.08^\circ$. The potential is in units of meV per graphene unit cell (UC).
\label{fig1}}
\end{figure*}

Here, full relaxation of the commensurate double moir\'{e} superlattices in TBG/BN with the twist angle around $\theta_m$
is performed and the single-particle
effective Hamiltonian ($\hat{H}^0$) with the relaxation effect and the full moir\'{e} Hamiltonian induced by BN taken into account
is obtained, based on which the mean-field (MF) band structures of the SCHF ground states are acquired in the plane-wave-like basis.
We find that the symmetry breaking in $\hat{H}^0$ mainly originates from the broken
$C_2$ symmetry in the structural deformation of the graphene layer adjacent to BN,
which introduces spatially non-uniform modifications of the intralayer Hamiltonian.
The gapped flat bands of $\hat{H}^0$ have finite Chern numbers.
For TBG/BN with $\theta_m$, the SCHF ground states with $|\nu|$ = 1-3 are all insulating
with narrow MF gaps.

The outline of this paper is as follows: In Sec. II we present the geometry and the structural
relaxation of the double moir\'{e} superlattices in TBG/BN.
For the relaxed TBG/BN, the symmetry breaking in $\hat{H}^0$ is studied in Sec. III and
the MF electronic structure at integer band filling is shown in Sec. IV.
Section V presents the summary and conclusions.

\section{Relaxation of the double moir\'{e} superlattices}

Placing TBG on a BN monolayer with their orientations nearly aligned, the double moir\'{e} superlattices
emerge in the TBG/BN trilayer, as depicted in Figs. 1(a) and 1(b).
We consider the TBG/BN with the twist angle ($\theta$) between the top graphene layer (G3)
and the fixed middle graphene layer (G2) close to $\theta_m$.

The spanning vectors of the moir\'{e} superlattice in G3 on G2 (G3/G2)
are along the armchair directions of graphene, while those of
the moir\'{e} superlattice in G2 on BN (G2/BN) with perfect alignment are along the zigzag directions.
The lattice-constant mismatch between G2 and BN and also the relative twist give rise to the
moir\'{e} superlattice in G2/BN.
The calculated lattice constant of graphene ($a$) is 2.447 {\AA}
and that of the BN monolayer ($a'$) is 2.489 {\AA},
where $a = (1 + \epsilon) a'$ with $\epsilon = -1.70\%$.
Since the double moir\'{e} superlattices have similar sizes but distinct orientations when TBG with $\theta_m$ is perfectly aligned with
BN, these superlattices are incommensurate. To obtain the commensurate double superlattices, the BN layer has to be rotated by
a small angle ($\theta'$) with respect to the fixed G2 layer.
We find that when $\theta'$ is around 1.6$^\circ$ and the lattice-constant mismatch between graphene and BN is maintained around
-1.70\%, the superlattices become commensurate and can be strictly periodic.
The size of the moir\'{e} supercell
in G2/BN is just a third of that in G3/G2, as indicated in Fig. 1(b).
Such strictly commensurate double moir\'{e} superlattices with tuned geometry parameters allow us to theoretically
study the electronic structure of fully relaxed TBG/BN in the plane-wave-like basis.
The geometry of the double moir\'{e} superlattices in TBG/BN is detailed in the Appendix
and the geometry parameters of the considered commensurate TBG/BN are listed in Table I.

The rigid moir\'{e} superlattices undergo spontaneous in-plane relaxation due to the energy gain
from the larger domains of energetically favorable stacking configurations.
Each layer is also corrugated to reach the optimal interlayer distances
of the varying stacking configurations across the superlattices.
In the superlattice between graphene layers, the interlayer
interaction potential ($V$) has the lowest and the same value at the AB- and BA-stacked parts,
and its spatial variation is inversion symmetric with respect to the AA-stacked part at the origin,
as shown in Fig. 1(d). The symmetry of $V$ is due to the $C_2$ symmetry in the rigid structure of TBG
with respect to the vertical axis for the relative rotation of the graphene layers.
Within each graphene layer,  $C_2$ becomes the inversion symmetry with respect to the hexagon center at the origin.
Upon relaxation in the pristine TBG, the regions with AB-like and BA-like stackings increase in size and
the $C_2$ symmetry is maintained.
In contrast, the $C_2$ symmetry is absent in the BN layer and also in the superlattice of G2/BN.
The BA-stacked parts in G2/BN have a much lower interlayer interaction potential ($V'$) than that at the
AB-stacked parts, leading to an inversion-asymmetric distribution of $V'$, as shown in Fig. 1(c).
Then, only the regions with BA-like stackings tend to become larger upon relaxation in G2/BN.
Therefore, subject to both $V$ and $V'$, the $C_2$ symmetry in G2 layer
is broken upon relaxation, and the in-plane structural deformation in G2 becomes distinct from that
in the relaxed pristine TBG. We have performed the full relaxation of TBG/BN
employing the continuum elastic theory, as detailed in the Appendix.

\begin{figure}[t]
\includegraphics[width=1\columnwidth]{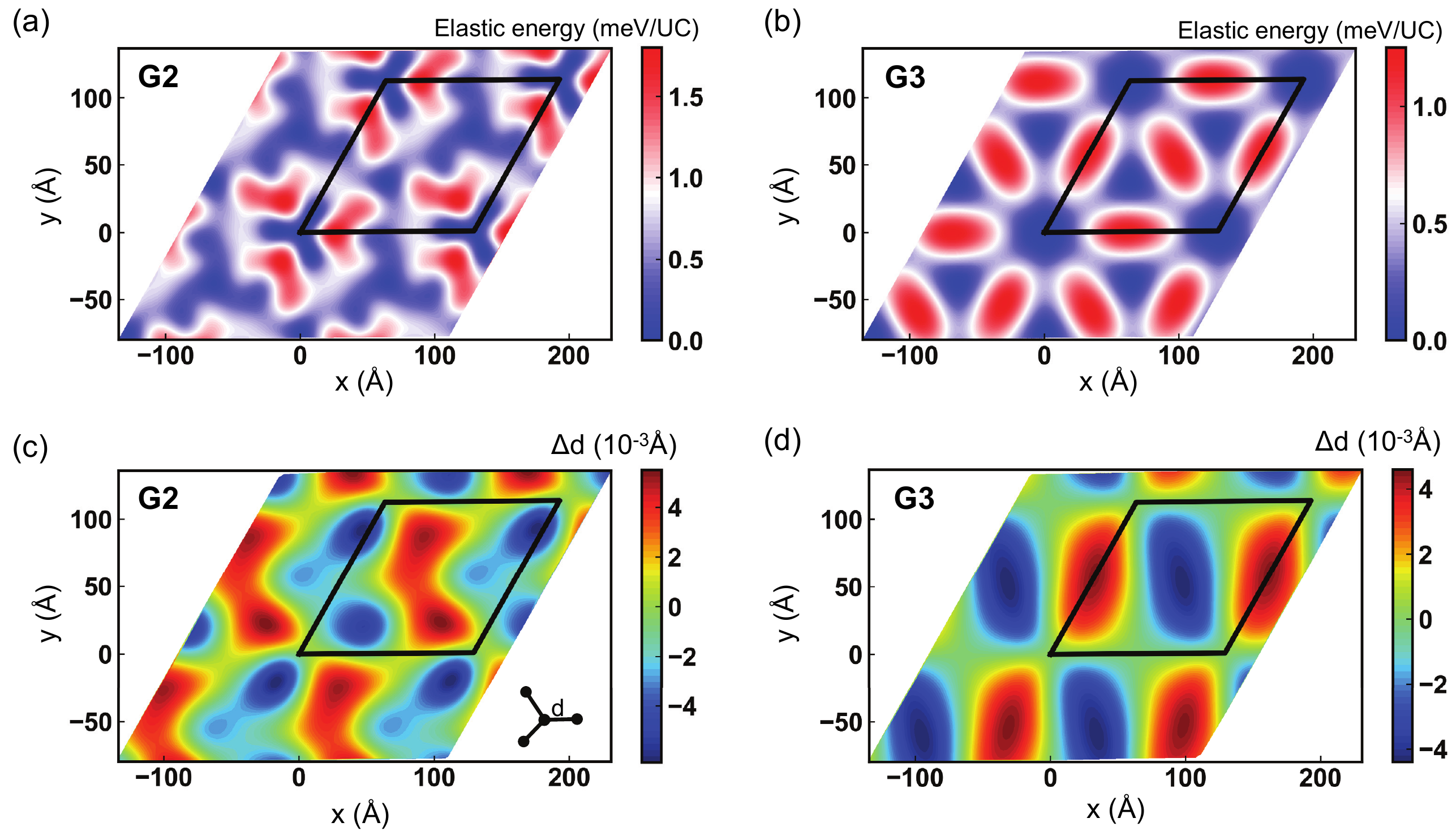}
\caption{(Color online) The in-plane structural deformation in the graphene layers of the relaxed
TBG/BN with $\theta = 1.08^\circ$.
(a) and (b) The spatial distribution of the local elastic energy in G2 (a) and G3 (b).
(c) and (d) The distribution of $\Delta d = d  - d_0$ in G2 (c) and G3 (d), where $d$ is the length of a C-C bond along the $x$ direction (see the inset)
and $d_0$ is the bond length of the rigid graphene.
An periodic supercell is indicated by the black lines in (a-d).
\label{fig2}}
\end{figure}

To demonstrate the broken symmetry in G2 of the relaxed TBG/BN and to compare the in-plane structural deformation in G2 and G3,
the spatial distributions of the elastic energy induced by
the strain field in the graphene layers are illustrated in Figs. 2(a) and 2(b).
Since the relaxation of G3 is driven by the interaction between graphene layers, the
spatial variation of the elastic energy in G3 is similar to that in the pristine TBG.
The regions with high elastic energy in G3 are located between the AB- and BA-stacked parts,
and the distribution is approximately inversion symmetric.
In contrast, the inversion symmetry is absent in the elastic-energy distribution of G2.
The regions in G2 between parts with BA stacking of G2/BN have higher elastic energy than the positions between
parts with AB and BA stackings of G3/G2.
In addition, the highest elastic energy in G2 is much larger than that in G3, indicating stronger relaxation
in G2.

The elastic-energy distribution reflects the spatial variation of the strain field in each graphene
layer.
As we will show in the next section, the strain field
induces changes in the intralayer on-site energies and hopping terms of the Hamiltonian.
The changing of the intralayer hopping terms is due to the modified C-C bond lengthes ($d$)
by strain. Figures 2(c) and 2(d) show the spatial variation of
$\Delta d = d - d_0$ for the  C-C bonds along the $x$ direction in G2 and G3,
respectively, where $d_0$ is the bond length of the rigid
graphene.
The bonds can be stretched or shortened by strain.
The regions with large $|\Delta d|$ in G2 are distinct from those in G3,
and the highest $|\Delta d|$ in G2 is larger than that in G3.
In particular, the distribution of $|\Delta d|$ in G2 has no inversion symmetry in contrast to that in G3.

\begin{table}[t]
\caption{The geometry parameters and electronic properties of
the studied commensurate double moir\'{e} superlattices in TBG/BN:
$N$ used to express the supercell basis vectors,
the twist angle ($\theta$) between G3 and G2, the lattice-constant mismatch between graphene and BN ($\epsilon$),
the twist angle ($\theta'$) between G2 and BN,
the widthes of the single-particle flat valence ($W_v$) and conduction ($W_c$) bands,
the single-particle gap ($\Delta_0$) at CNP, and the MF gaps
($\Delta_\nu^{\mathrm{MF}}$) in the SCHF ground states at different number ($\nu$) of filled flat bands.
$W_v$, $W_c$, $\Delta_0$, and $\Delta_\nu^{\mathrm{MF}}$ are in units of meV.
}
\begin{tabular*}{0.48\textwidth}{@{\extracolsep{\fill} }rrrrrrrrrrrrrrr}
\hline\hline
$N$ & 30  & 29 & 28 \\
\hline
$\theta$    & 1.0845$^\circ$ & 1.1213$^\circ$ &    1.1607$^\circ$\\
$\epsilon$  & -1.6255\% & -1.6801\% &    -1.7385\%\\
$\theta'$   & 1.6448$^\circ$ & 1.7012$^\circ$ &    1.7616$^\circ$\\
$W_v$   &  2.75   &  3.70  & 7.83  \\
$W_c$   &  2.50   &  7.23  & 12.27  \\
$\Delta_0$    &  6.92   &  7.85  &  8.64 \\
$\Delta_1^{\mathrm{MF}}$   &  3.76   &  13.41  &  19.49\\
$\Delta_2^{\mathrm{MF}}$   &  3.51   &  15.08  &  20.48 \\
$\Delta_3^{\mathrm{MF}}$   &  4.78   &  16.74  &  21.17 \\
$\Delta_{\text{-1}}^{\mathrm{MF}}$   &  5.76   &  10.36  &  17.73  \\
$\Delta_{\text{-2}}^{\mathrm{MF}}$   &  6.92   & 11.01   & 18.80  \\
$\Delta_{\text{-3}}^{\mathrm{MF}}$   &  7.63   & 12.59   & 17.12  \\
\hline\hline
\end{tabular*}
\end{table}

\section{Symmetry breaking in the single-particle Hamiltonian of relaxed TBG/BN}

For the relaxed TBG/BN, we have built an effective single-particle Hamiltonian $\hat{H}^0$ for the
moir\'{e} superlattice in G3/G2 by extending the Hamiltonian of $p_z$ orbitals for graphene bilayers
proposed in Refs.~[\onlinecite{ShearPhysRevB.98.195432}, \onlinecite{Pressure2020lin}]
and the effective Hamiltonian of monolayer graphene on BN proposed in Ref.~[\onlinecite{Effective2019lin}].
This effective Hamiltonian reads
\begin{eqnarray}
\hat{H}^0 &=& \sum_{n=2}^{3} \sum_{i} \varepsilon_{n, i} c^{\dagger}_{n,i} c_{n,i} +  \sum_{n=2}^{3} \sum_{\left<i,j\right>} t^{(n,n)}_{i,j} (c^{\dagger}_{n,i} c_{n,j} + h.c.)\nonumber \\
  &+& \sum_{i,j} t^{(2,3)}_{i,j} (c^{\dagger}_{2,i} c_{3,j} + h.c.),
\end{eqnarray}
where $c^{\dagger}_{n,i}$ $(n=2,3)$ is the creation and $c_{n,i}$ is the annihilation operator of
a $p_z$-like orbital at the site $i$ in the G$n$ layer,
$\left<i,j\right>$ denotes the intralayer nearest neighbors, and
the on-site energies, intralayer and interlayer hopping terms
are represented by $\varepsilon_{n, i}$, $t^{(n,n)}_{i,j}$, and $t^{(2,3)}_{i,j}$, respectively.
The used parameters to obtain these on-site and hopping terms are extracted from the \emph{ab-initio} electronic
structures of shifted bilayers.

\begin{figure*}[t]
\includegraphics[width=1.7\columnwidth]{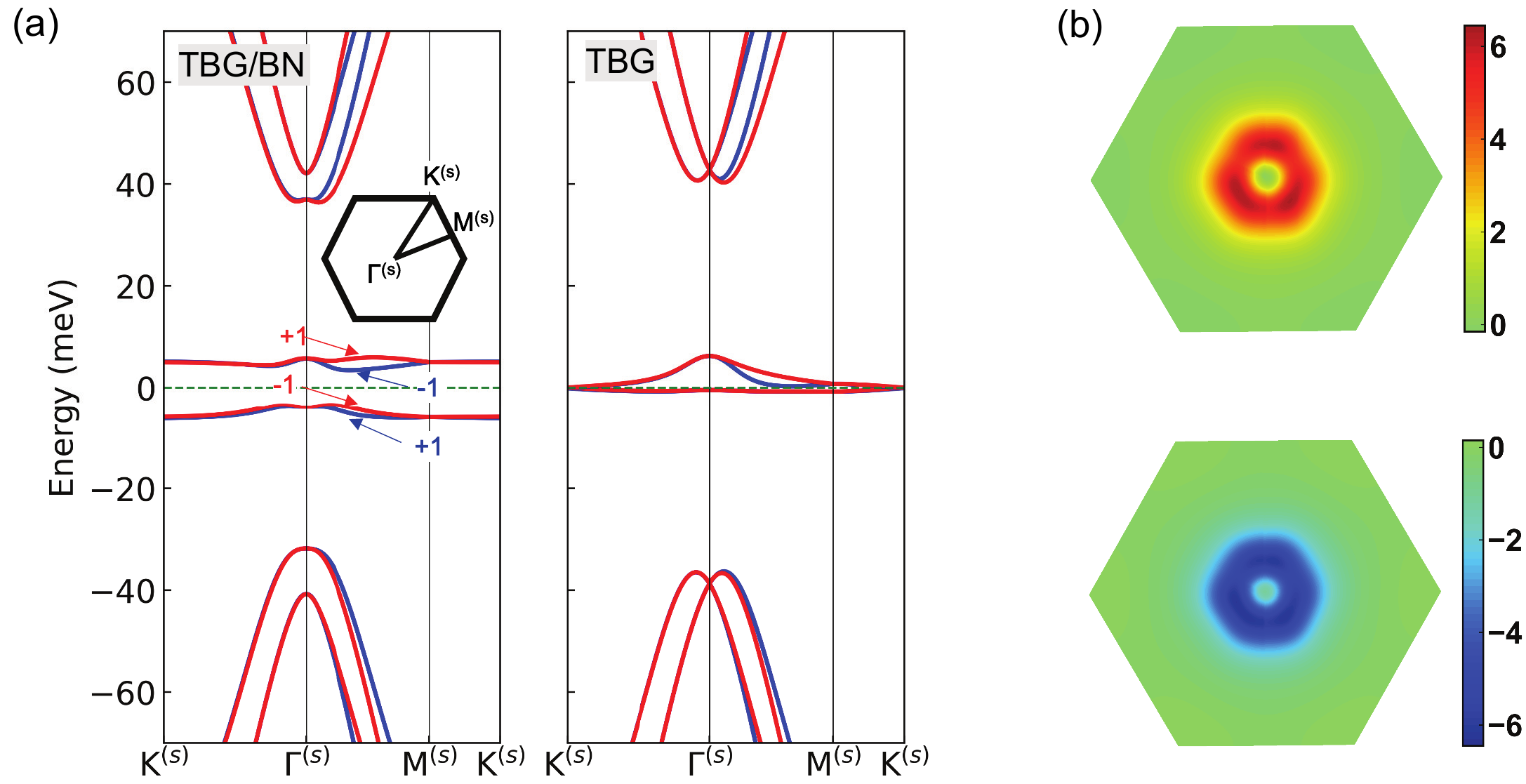}
\caption{(Color online) (a) The single-particle band structures of the relaxed TBG/BN and
pristine TBG with $\theta = 1.08^\circ$.
The blue and red lines
represent bands in the $\xi = +$ and $\xi = -$ valleys, respectively.
The Fermi levels
are set to be zero and are represented by the dashed green lines.
The Chern number of each flat band is labeled.
(b) The distribution of the Berry curvature in units of {\AA}$^2$ in the supercell BZ for the flat valence (top panel) and conduction (bottom panel) bands
in the $\xi = +$ valley.
\label{fig3}}
\end{figure*}

For the intralayer on-site and hopping terms, we take into account the effect of BN and the in-plane strain.
The strain modifies the bond lengthes in graphene and thus influences the
intralayer hopping terms.
The hopping $t^{(n,n)}_{i,j}$ between intralayer nearest neighbors with their distances $d$ deviated from that of
the pristine graphene $d_0 = a/\sqrt{3}$ is given by
\begin{equation}
V_{pp\pi}(d, \bm{r}) = -V_{pp\pi}^0(\bm{r}) e^{-(d - d_0)/\lambda_{\pi}}
\end{equation}
with $\lambda_{\pi} = 0.47$ {\AA}.
The $V_{pp\pi}^0$ for G3 is a constant and is taken to be 2.598 eV, while
the $V_{pp\pi}^0$ for G2 depends on the local shift vector $\bm{\delta}'(\bm{r})$ between G2 and BN
with $V_{pp\pi}^0(\bm{\delta}')$ given by the effective model in Ref.~[\onlinecite{Effective2019lin}].
In G2, the on-site energies in the two sublattices ($\varepsilon_A$ and $\varepsilon_B$) also vary with
$\bm{\delta}'$ and can be computed by the model in Ref.~[\onlinecite{Effective2019lin}].
Moreover, the in-plane strain induces changes in $\varepsilon_m$ ($m$ = A, B) for both G2 and G3
expressed as
\begin{equation}
\delta \varepsilon_m = \alpha_0 \left(\frac{\partial u_x^{(n)}}{\partial x} + \frac{\partial u_y^{(n)}}{\partial y}\right),
\end{equation}
where $n$ = 2, 3, and $\alpha_0 = -4.95$ eV.
Calculations show that the $\delta \varepsilon_i$ in G3 is smaller than 1 meV, while it can be rather large in G2.

The interlayer hopping $t^{(2,3)}_{i,j}$ between
sites in G2 and G3 with in-plane projection $r$ and out-of-plane projection $h$ is expressed as
\begin{equation}
V_{pp\sigma}(r, h) = V_{pp\sigma}^0 e^{-(h - h_0)/\lambda'} e^{-(\sqrt{r^2 + h^2} - h)/\lambda} \frac{h^2}{r^2 + h^2},
\end{equation}
where $V_{pp\sigma}^0$ = 0.381 eV, $h_0$ = 3.32 {\AA}, $\lambda'$ = 0.58 {\AA}, and $\lambda$ = 0.27 {\AA}.
All interlayer hopping terms with $r \leq 5.0$ {\AA} are included in the calculations.
The local optimal interlayer spacing obtained from \emph{ab-initio} calculations is given by
$h(\bm{\delta}) = 3.413 + 0.0622 \sum_{j=1}^{3} \cos(\mb{G}_{j} \cdot \bm{\delta})$
in units of {\AA}.
We note that these Hamiltonian parameters can reproduce the observed $\theta_m$ of the pristine
TBG.

Since the moir\'{e} supercell in G3/G2 is rather large, the Hamiltonian $\hat{H}^0$ can be
expressed using the plane-wave-like basis functions.
In this approach, the atomic positions of the rigid
graphene lattice in each layer are used to label the hopping sites in the Hamiltonian.
The plane-wave-like basis functions are labeled with the valley index ($\xi = \pm$),
the sublattice and layer index ($\alpha$ = A2, B2, A3, B3), a k-point ($\mb{k}^{(s)}$) in the supercell BZ,
and a reciprocal lattice vector (${\bf{G}^{(s)}}$) of the
superlattice. They are defined as
\begin{eqnarray}
|\alpha, \mb{k}^{(s)} &+& \mb{k}_{\xi} + \mb{G}^{(s)}\rangle =  \nonumber \\
& &\frac{1}{\sqrt{\tilde{N}}} \sum_{\mb{r}_{\alpha}}
e^{i (\mb{k}^{(s)} + \mb{k}_{\xi} + \mb{G}^{(s)}) \cdot \mb{r}_{\alpha}}
|\mb{r}_{\alpha}\rangle,
\end{eqnarray}
where $\mb{k}_{\xi}$ is the center
of one of the supercell BZs containing the Dirac points of each layer at their
corners in the $\xi$ valley and $\mb{r}_{\alpha}$ is the rigid in-plane position of a site in the sublattice $\alpha$ of the corresponding layer.
$\mb{k}_{\xi}$ is thus a reciprocal lattice vector of the supercell and
the used $\mb{k}_{\xi}$ can be seen in the schematic reciprocal lattice of a moir\'{e} superlattice in Fig. 7(a) of the Appendix.
We use the 37 shortest ${\bf{G}^{(s)}}$ shown in Fig. 7(b).
The $\hat{H}^0$ element between two basis functions is given by
\begin{eqnarray}
H^0_{\alpha,\mb{G}^{(s)'};\beta,\mb{G}^{(s)}}(\xi,\mb{k}^{(s)}) =
\langle \alpha, \mb{k}' | \hat{H}^0 |\beta, \mb{k} \rangle = \nonumber \\
\frac{1}{\tilde{N}_0}  \sum_{\mb{r}_{\alpha} \in SC} \sum_{\mb{r}_{\beta}}
e^{-i \mb{k}' \cdot \mb{r}_{\alpha} + i \mb{k} \cdot \mb{r}_{\beta}}
\langle \mb{r}_{\alpha} |\hat{H}^0| \mb{r}_{\beta} \rangle,
\end{eqnarray}
where $\mb{k}' = \mb{k}^{(s)} + \mb{k}_\xi+\mb{G}^{(s)'}$,
$\mb{k} = \mb{k}^{(s)} + \mb{k}_\xi+\mb{G}^{(s)}$,
the summation over $\mb{r}_{\alpha}$ is done in a supercell, and $\tilde{N}_0$ is the number of graphene unit cells in one layer of the supercell.
$\langle \mb{r}_{\alpha} |\hat{H}^0| \mb{r}_{\beta} \rangle$ represents the on-site and hopping terms given above.
For each $\mb{r}_{\alpha}$, only a small number of large $\langle \mb{r}_{\alpha} |\hat{H}^0| \mb{r}_{\beta} \rangle$ are
required in the summation of Eq. (6).
Since the Hamiltonian between states from two different valleys is negligible for large
moir\'{e} superlattices, the bands from $H^0$ are valley polarized.
The Berry curvature of the band states and the Chern number of each band can be calculated for the Hamiltonian in Eq. (6) using
the method in Ref. [\onlinecite{Pressure2020lin}].

\begin{figure*}[t]
\includegraphics[width=1.6\columnwidth]{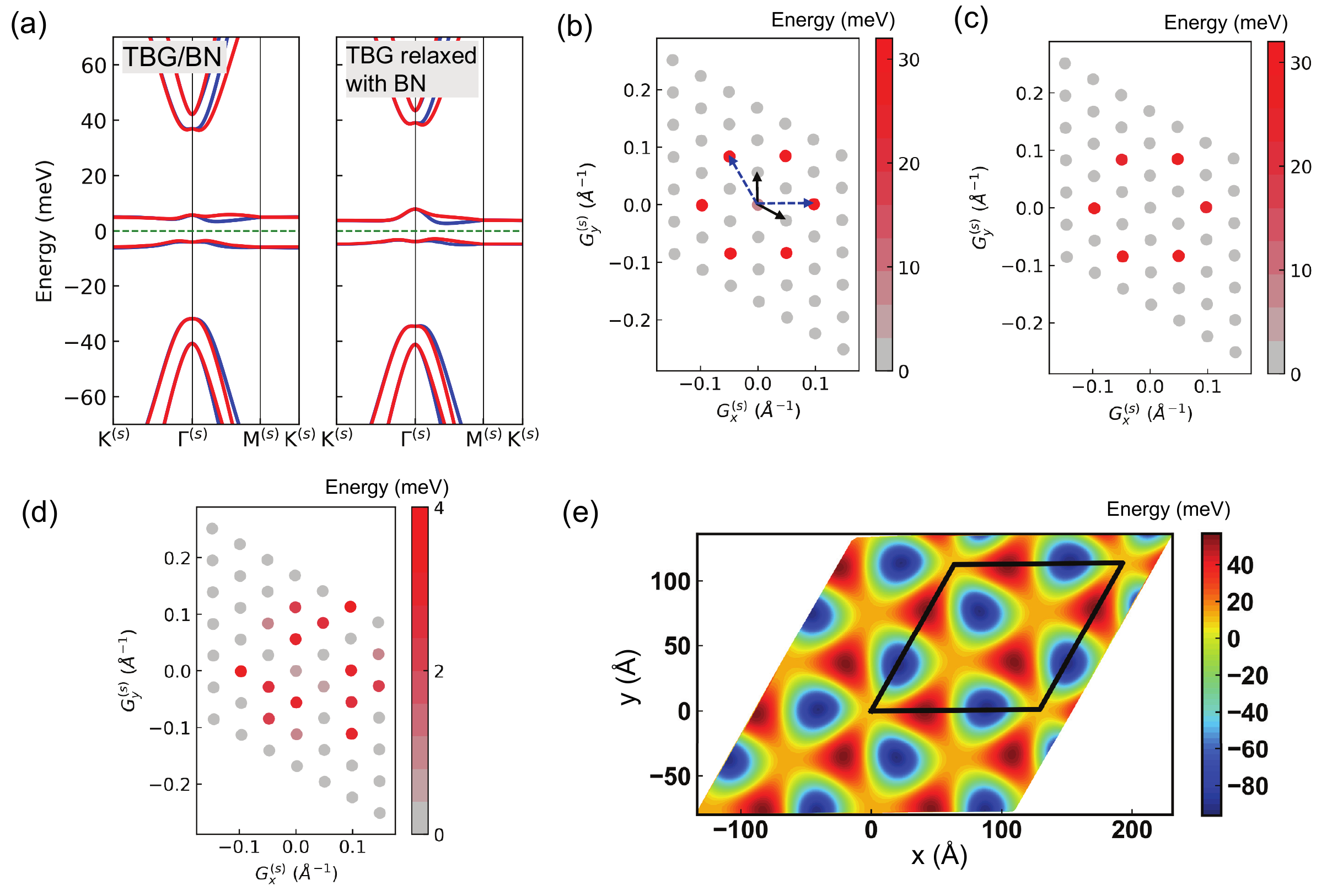}
\caption{(Color online) (a) The single-particle band structure of the relaxed TBG/BN with $\theta = 1.08^\circ$ compared with that of G3/G2
ignoring the effective Hamiltonian in G2 introduced by BN while still considering the interlayer potential between G2 and BN
for the structural relaxation.
(b)-(d) The differences between the $H^0$ elements due to the broken $C_2$ symmetry as a function of $\mb{G}^{(s)}$
for the $\xi = +$ valley and the $K^{(s)}$ point:
$|H^0_{A2,\mb{0};A2,\mb{G}^{(s)}} - H^{0*}_{B2,\mb{0};B2,\mb{G}^{(s)}}|$ (b),
$|H^0_{A2,\mb{0};B2,\mb{G}^{(s)}} - H^{0*}_{B2,\mb{0};A2,\mb{G}^{(s)}}|$ (c), and
$|H^0_{A2,\mb{0};A3,\mb{G}^{(s)}} - H^{0*}_{B2,\mb{0};B3,\mb{G}^{(s)}}|$ (d).
In (b) the solid and dashed lines represent the spanning vectors of the reciprocal lattices of
the moir\'{e} superlattices in G3/G2 and G2/BN, respectively.
(e) The distribution of the average on-site energies $(\varepsilon_A+\varepsilon_B)/2$ in G2.
\label{fig4}}
\end{figure*}

The single-particle band structure of the relaxed TBG/BN with $\theta = 1.08^\circ$ is displayed in Fig. 3(a).
In contrast to the pristine TBG with degenerate conduction and valence band states at $K^{(s)}$,
the flat bands of TBG/BN are separated by the opened gap. This is due to
the broken $C_2$ symmetry in $\hat{H}^0$ by both relaxation and BN.
For TBG/BN with $\theta = 1.08^\circ$, the gap ($\Delta_0$) at CNP reaches 6.92 meV.
The flat conduction and valence bands in TBG/BN have similar band widthes of about 2.5 meV, while the flat valence bands
in the pristine TBG are much narrower than the flat conduction bands.
In both TBG/BN and the pristine TBG, the flat bands are well gapped from other dispersive bands.
Since the flat valence and conduction bands have become separated in TBG/BN, the Chern number ($C_n$) of each flat band can be computed to
characterize its topological property.
In the $\xi = +$ valley, the flat valence and conduction bands have finite $C_n$ of $+1$ and $-1$, respectively.
The $C_{n}$ of each flat band in the $\xi = -$ valley is just
the opposite of that for $\xi = +$, as labeled in Fig. 3(a).
The finite $C_n$ of the flat bands are contributed by the large Berry curvature at k-points around the center of the supercell Brillouin zone (BZ), as shown in Fig. 3(b).

The main characteristics of the electronic structure of relaxed TBG/BN in contrast to the pristine TBG
are caused by the broken $C_2$ symmetry in $\hat{H}^0$. We will reveal the origin of
this symmetry breaking in the following.
The BN layer introduces an effective Hamiltonian in G2, whose on-site and hopping terms lack the $C_2$ symmetry.
The $C_2$ symmetry is present in the atomic structure of the pristine TBG,
while it is broken in the relaxed structure of G3/G2 in TBG/BN.
The structural deformation without the $C_2$ symmetry also induces changes in the intralayer on-site and hopping terms as well as the
interlayer hopping terms compared with those of the pristine TBG.
In Fig. 4(a), the single-particle band structure of the relaxed TBG/BN is compared with that of G3/G2
ignoring the effective Hamiltonian in G2 introduced by BN while still considering the interlayer potential between G2 and BN
for the structural relaxation. We find that the flat bands have similar dispersions, and the gap $\Delta_0$ is just reduced by
1.3 meV when the electronic contribution of BN is ignored. This indicates that the symmetry breaking in $\hat{H}^0$ mainly originates from
the broken $C_2$ symmetry in the relaxed structure of G3/G2 in TBG/BN.

For the pristine TBG, $\hat{H}^0$ remains the same under the operation of $C_2$, that is $\hat{H}^0 = C_2 \hat{H}^0 C_2$.
In addition, the $C_2$ operation transforms the plane-wave-like basis functions defined in Eq. (5) as
$C_2 |\tilde{\alpha}n, \mb{k}\rangle = |(-\tilde{\alpha})n, -\mb{k}\rangle$, where $n$ is the layer index and $\tilde{\alpha} = +$
and $\tilde{\alpha} = -$ denote the A and B sublattices, respectively.
Considering both the $C_2$ and time reversal ($\mathcal{T}$) symmetries,
the elements of $\hat{H}^0$ for the pristine TBG satisfy $\langle \tilde{\alpha}'n', \mb{k}' | \hat{H}^0 |\tilde{\alpha}n, \mb{k} \rangle
= \langle (-\tilde{\alpha}')n', \mb{k}' | \hat{H}^0 |(-\tilde{\alpha})n, \mb{k} \rangle^*$.
Then, the magnitude of the difference $|\langle \tilde{\alpha}'n', \mb{k}' | \hat{H}^0 |\tilde{\alpha}n, \mb{k} \rangle
- \langle (-\tilde{\alpha}')n', \mb{k}' | \hat{H}^0 |(-\tilde{\alpha})n, \mb{k} \rangle^*|$ can be used to indicate the
strength of $\hat{H}^0$ terms associated with the broken $C_2$ symmetry in TBG/BN.

We take the $H_0$ at $K^{(s)}$ in the $\xi = +$ valley as an example
and denote $\langle \tilde{\alpha}'n', \mb{k}' | \hat{H}^0 |\tilde{\alpha}n, \mb{k} \rangle$
with $\mb{k}' = K^{(s)} + \mb{k}_+ +\mb{G}^{(s)'}$ and
$\mb{k} = K^{(s)} + \mb{k}_++\mb{G}^{(s)}$ by $H^0_{\tilde{\alpha}'n',\mb{G}^{(s)'};\tilde{\alpha}n,\mb{G}^{(s)}}$.
The elements $H^0_{\tilde{\alpha}'n',\mb{0};\tilde{\alpha}n,\mb{G}^{(s)}}$ with $\mb{G}^{(s)'} = \mb{0}$ are considered.
Their corresponding $|H^0_{\tilde{\alpha}'n',\mb{0};\tilde{\alpha}n,\mb{G}^{(s)}} -
H^0_{(-\tilde{\alpha}')n',\mb{0};(-\tilde{\alpha})n,\mb{G}^{(s)}}|$ as a function of
$\mb{G}^{(s)}$ are illustrated in Figs. 4(b)-4(d).
The $H^0_{A2,\mb{0};A2,\mb{G}^{(s)}}$ and $H^{0}_{B2,\mb{0};B2,\mb{G}^{(s)}}$ are contributed by
the on-site energies $\varepsilon_A$ and $\varepsilon_B$ in G2, which are mainly induced by
the strain field [see Eq. (3)].
We find that $|H^0_{A2,\mb{0};A2,\mb{G}^{(s)}} - H^{0*}_{B2,\mb{0};B2,\mb{G}^{(s)}}|$
has large values only at the shortest nonzero reciprocal lattice vectors of the moir\'{e} superlattice in
G2/BN, and the value at zero $\mb{G}^{(s)}$ is finite but small.
This implies that the symmetry breaking in the on-site terms of $\hat{H}^0$ is mainly due to the nonuniform
inversion-asymmetric spatial variation of $\varepsilon_A$ and $\varepsilon_B$ in G2 [see Fig. 4(e)] rather than the
uniform $\varepsilon_A - \varepsilon_B$.
Such distribution of on-site energies in G2 is just due to the absence of inversion symmetry in the strain field
shown in Fig. 2(a). In contrast, the on-site energies in G3 are approximately symmetric but are all smaller than 1 meV.

The intralayer hopping between nearest neighbors changes when the bond lengthes $d$
deviate from $d_0$ of the pristine graphene [see Eq. (2)].
In G3, the distribution of $\Delta d = d - d_0$ is approximately inversion symmetric, as seen in Fig. 2(c), the symmetry is thus
maintained for the hopping terms in G3.
However, the inversion-asymmetric $\Delta d$ in G2 [see Fig. 2(d)] leads to large
$|H^0_{A2,\mb{0};B2,\mb{G}^{(s)}} - H^{0*}_{B2,\mb{0};A2,\mb{G}^{(s)}}|$, which is contributed by the intralayer hopping in G2,
as shown in Fig. 4(c). The locations with high values in Fig. 4(c) are the same as those in Fig. 4(b) and the magnitudes in both
figures are similar.
The symmetry of the interlayer hopping is also affected by the structural deformation in G2, while the influence is much weaker, as
indicated by the small $|H^0_{A2,\mb{0};A3,\mb{G}^{(s)}} - H^{0*}_{B2,\mb{0};B3,\mb{G}^{(s)}}|$ shown in Fig. 4(d).

\section{Mean-field band structures at integer band filling}

When an integer number more or less flat bands are filled relative to CNP in relaxed TBG/BN,
electron-electron (e-e) interaction has to be taken into account explicitly to describe the insulating states
at $|\nu|=1-3$ observed in experiments.
Due to the large moir\'{e} supercell, the envelop of a low-energy plane-wave-like basis function defined in Eq. (5)
varys slowly across the supercell and can be approximated by a continuous plane-wave function. Then
the e-e interaction can be expressed in the plane-wave-like basis as\cite{Spontaneous2019liu}
\begin{eqnarray}
\hat{H}_{e-e} &=& \frac{1}{2\tilde{N}}
  \sum_{\sigma\sigma'}\sum_{\xi\xi'} \sum_{\alpha\beta} \sum_{\mb{k}\mb{k'}\mb{q}} V_{\alpha\beta}(\mb{q}) \nonumber \\
  && c^{\dag}_{\mb{k}+\mb{q},\alpha\xi\sigma} c^{\dag}_{\mb{k'}-\mb{q},\beta\xi'\sigma'}
  c_{\mb{k'},\beta\xi'\sigma'} c_{\mb{k},\alpha\xi\sigma}\,,
\end{eqnarray}
where $\sigma$ and $\sigma'$ are the spin indexes, $V_{\alpha\beta}(\mb{q})$ is the e-e interaction kernel,
$\mb{k} = \mb{k}^{(s)} +\mb{G}^{(s)}$ and $\mb{k'} = \mb{k'}^{(s)} +\mb{G'}^{(s)}$ represent k-points with respect to $\mb{k}_\xi$ and
$\mb{k}_{\xi'}$, respectively,
$\mb{k}$ and $\mb{k}+\mb{q}$ with a small $\mb{q}$ are in the same valley and the terms with $\mb{k}$ and $\mb{k}+\mb{q}$
in different valleys are negligible.
The $V_{\alpha\beta}(\mb{q})$ is given by
$e^2/(\Omega^{(s)}4\pi \epsilon_r \epsilon_0)  2\pi/|\mb{q}|$
for sublattices  $\alpha$ and $\beta$ in the same layer and
$e^2/(\Omega^{(s)}4\pi \epsilon_r \epsilon_0)  e^{-|\mb{q}| \bar{h}} 2\pi /|\mb{q}|$ for
$\alpha$ and $\beta$ in different layers, where $\Omega^{(s)}$ is the supercell area,
$\epsilon_r$ is the dielectric constant, and $\bar{h}$ is the average interlayer spacing\cite{Nature2020xie}.
Here, $\epsilon_r$ is taken to be 10 considering the screening by both the BN substrates and electric gates.
With this $\epsilon_r$, the screened e-e interaction between intralayer nearest neighbors is 1.02 eV,
which is much smaller than that of the freestanding graphene without dielectric screening
from the substrates and electric gates\cite{Strength2011}.

\begin{figure}[t]
\includegraphics[width=1\columnwidth]{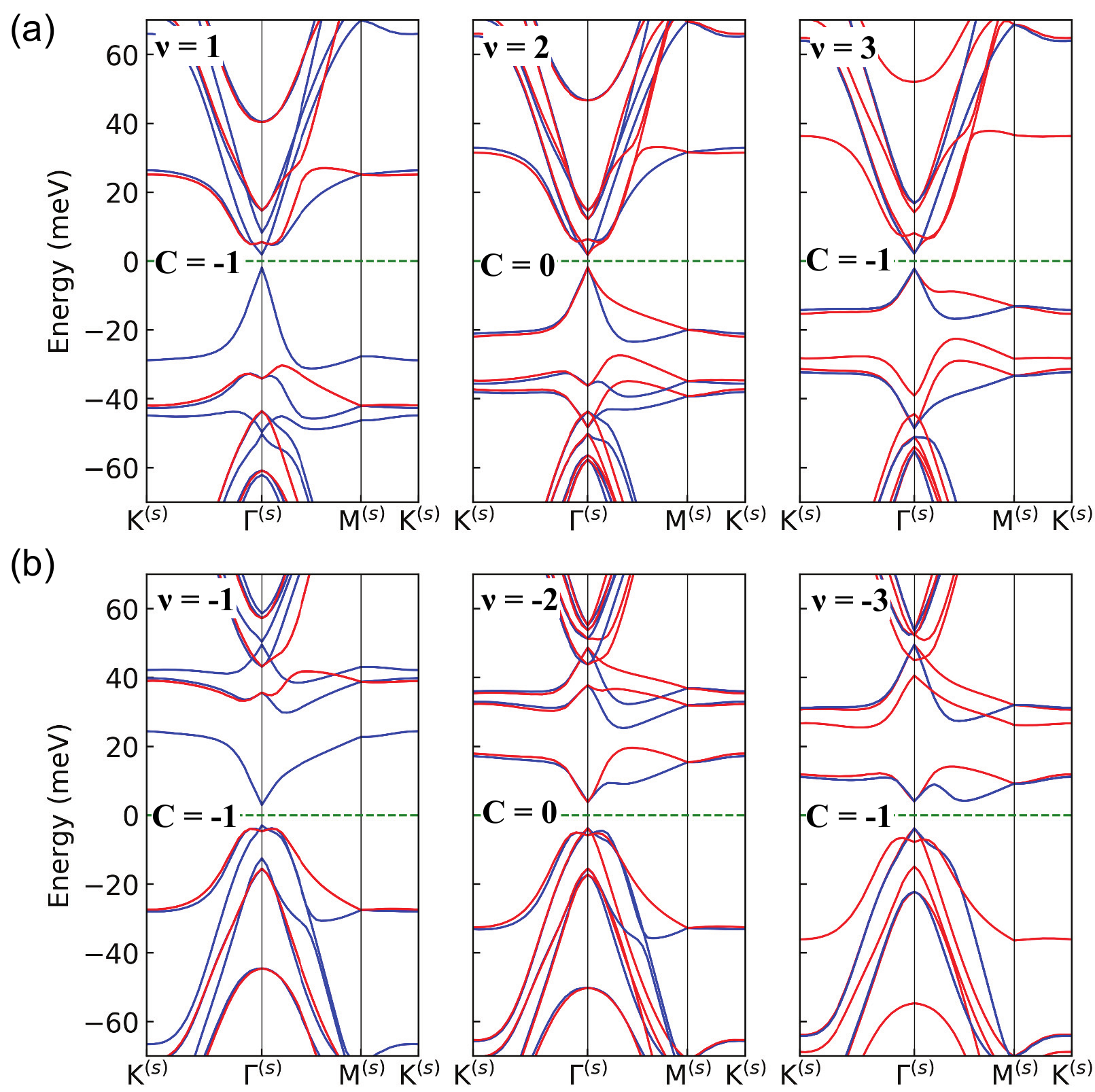}
\caption{(Color online) The MF band structures of the relaxed TBG/BN with $\theta = 1.08^\circ$
and an integer number of filled flat bands
in the SCHF ground states.
Positive (a) and negative (b) $\nu$ represent the number of filled conduction bands and the number of empty
valence bands, respectively.
Relative to CNP, one conduction band per valley is filled at $\nu = 2$,
two (one) conduction bands in the $\xi = +$ ($\xi = -$) valley are filled  at $\nu = 3$,
one valence band per valley is empty at $\nu = -2$, and two (one) valence bands in the $\xi = +$ ($\xi = -$) valley are empty  at $\nu = -3$.
The Fermi levels
are set to be zero and are represented by the dashed green lines.
The total Chern number ($C$) of the filled bands is labeled for each $\nu$.
\label{fig5}}
\end{figure}

We employ the SCHF method similar to that proposed by Xie and MacDonald\cite{Nature2020xie}
to obtain the mean-field (MF) band structures
of TBG/BN when the e-e interaction is taken into account.
It has been demonstrated that the insulating ground states of the pristine TBG at finite $\nu$
could be properly described by the SCHF method\cite{Nature2020xie}.
It is noted that expressing $\hat{H}^0$ in the plane-wave-like basis enables us
to obtain the SCHF ground states of the fully relaxed TBG/BN, while previous studies only considered rigid TBG or some relaxation effect in TBG with
one empirical parameter\cite{Nature2020xie,zhang2020correlated}.
The MF states with valley- and spin-polarized bands are considered here.
In the following, all k-points are for the supercell and the superscript `` (s)" in the k-point denotations is omitted
for clarity, and $\tau = (\xi, \sigma)$ is used to denote the four flavors of the band states.

Within the SCHF method, the MF Hamiltonian can be written as
$\hat{H}_{MF} = \hat{H}^0 + \hat{H}^H + \hat{H}^F$.
In the plane-wave-like basis, the single-particle Hamiltonian operator reads
$\hat{H}^0 = 1/N_{\mb{k}} \sum_{\sigma\xi}\sum_{\alpha\beta} \sum_{\mb{k}\mb{G}\mb{G'}}
H^0_{\alpha,\mb{G'};\beta,\mb{G}}(\xi,\mb{k}) c^{\dag}_{\mb{k}+\mb{G'},\alpha\xi\sigma}c_{\mb{k}+\mb{G},\beta\xi\sigma}$,
where $H^0_{\alpha,\mb{G'};\beta,\mb{G}}(\xi,\mb{k})$ is just the matrix element in Eq. (6).
The nonlocal Fock and Hartree operators are given by
\begin{eqnarray}
\hat{H}^F &=& -\frac{1}{N_{\mb{k}}} \sum_{\tau\mb{k}} \sum_{\alpha\beta\mb{G}\mb{Q}} \sum_{\mb{k}'\mb{G}'}  V_{\alpha\beta}(\mb{k} + \mb{G} + \mb{Q} - \mb{k}' - \mb{G}') \nonumber \\
       && \delta\rho_{\beta,\mb{G}'-\mb{Q}; \alpha,\mb{G}'}(\tau, \mb{k}') c^{\dag}_{\mb{k}+\mb{G}+\mb{Q},\alpha\tau} c_{\mb{k}+\mb{G},\beta\tau}\,, \\
\hat{H}^H &=& \frac{1}{N_{\mb{k}}} \sum_{\tau\mb{k}} \sum_{\alpha\mb{G}\mb{Q}} \sum_{\tau'\alpha'\mb{k}'\mb{G}'}  V_{\alpha\alpha'}(\mb{Q}) \nonumber \\
  &&   \delta\rho_{\alpha',\mb{G}'-\mb{Q}; \alpha',\mb{G}'}(\tau', \mb{k}') c^{\dag}_{\mb{k}+\mb{G}+\mb{Q},\alpha\tau} c_{\mb{k}+\mb{G},\alpha\tau}\,,
\end{eqnarray}
where $\delta\rho = \rho - \rho_{iso}$ with
$\rho_{\beta,\mb{G}'-\mb{Q}; \alpha,\mb{G}'}(\tau, \mb{k}')$ denoting the density matrix element
$\left<c^{\dag}_{\mb{k'}+\mb{G'}-\mb{Q},\beta\tau}c_{\mb{k'}+\mb{G'},\alpha\tau}\right>$ for the MF state and
$\rho_{iso}$ the density matrix for the isolated fixed and rotated graphene layers with band states below the
charge neutrality point occupied, and
the summation over $\mb{k}$ and $\mb{k}'$ is done in the supercell BZ with
a uniform $30 \times 30$ k-point grid.
$\rho$ can be calculated as
\begin{equation}
\rho_{\beta,\mb{G}'-\mb{Q}; \alpha,\mb{G}'}(\tau, \mb{k}') = \sum_{n}^{\text{occupied}}\psi^{n*}_{\beta,\mb{G}'-\mb{Q}}(\tau, \mb{k}') \psi^{n}_{\alpha,\mb{G}'}(\tau, \mb{k}'),
\end{equation}
where $\psi^{n}(\tau, \mb{k}')$ represents a MF band state with flavor $\tau$ at the k-point $\mb{k}'$.
In Eqs. (8) and (9), 19 shortest reciprocal lattice vectors [see Fig. 7(b)] are used for $\mb{Q}$ as the
Fock and Hartree elements generally decease with
the magnitude of $\mb{Q}$.
With such a set of $\mb{Q}$, the non-locality of the Fock and Hartree operators is maintained in the calculations.
The summation over terms involving $V_{\alpha\beta}(\mb{q})$ with $\mb{q}$ close to zero in Eq. (8) is done
using a similar method in Ref. [\onlinecite{Self1986}].

The $\hat{H}_{MF}$ can be diagonalized to obtain the MF band states for each $\tau$ and $\mb{k}$.
We have used the 37 shortest $\mb{G}$ shown in Fig. 7(b) for the plane-wave-like basis.
Then 148 bands per flavor are updated during the SCHF iterations.
The MF Hamiltonian and its total energy are functionals of the density matrix $\rho$.
For a filling factor $\nu$, the initial $\rho$ of the self-consistent calculation is taken to be
that of the single-particle band states with the filling scheme corresponding to $\nu$.
Then a linearly mixed $\rho$ is used for the next iteration.
The SCHF iterations are performed until the convergence of the total energy per supercell is reached with a tolerance of
$10^{-5}$ eV.
The filling scheme is determined by the band-state energies during the iterations, and we find that
it remains the same as the initial one for the converged state.

Figure 5 shows the computed MF band structures
at $\nu = \pm1$, $\pm2$, and $\pm3$ for TBG/BN with $\theta = 1.08^\circ$.
The SCHF ground states at different $\nu$ are all insulating with narrow MF gaps.
The total Chern number ($C$) of the filled bands is also calculated and are labeled in Fig. 5.
At $\nu = \pm1$ and $\pm3$, a nontrivial $C$ is obtained and the QAH effect may be observed.
For $\nu = +1$ and $+3$, one more flat band with $C_n = -1$ in the $\xi = +$ valley is occupied than the
$\xi = -$ valley, and one more flat band with $C_n = +1$ in the $\xi = +$ valley is empty
than the $\xi = -$ valley for $\nu = -1$ and $-3$.

The MF gaps opened around the $\Gamma^{(s)}$ point are listed in Table I.
When the flat conduction bands are filled, the gap at $\nu$ = 1 is smaller than that at $\nu$ = 3,
the QAH effect is thus more likely to be observed at $\nu = 3$.
This is roughly consistent with the recent experimental observations\cite{Intrinsic2020Serlin}.
For states with empty valence bands, the gap at $\nu$ = -3 is also larger than that $\nu$ = -1, while
their gaps are larger than those at positive $\nu$.
Compared with the experiments\cite{Intrinsic2020Serlin}, the band gap at $\nu = 3$ is overestimated at the MF level,
which may be corrected by a higher theoretical level with the dynamically screened e-e interaction in future investigations.
For TBG/BN with larger $\theta$, the gaps become wider and their values at negative $\nu$
become smaller than those at positive $\nu$, as shown in Table I.

To characterize the nature of the MF band states, the local density of states (LDOS) at the AA-stacked part of
G3/G2 is computed for each $\nu$ and is compared with that of the single-particle bands at CNP, as shown in Fig. 6.
The single-particle flat-band states are mainly localized at the AA-stacked part as demonstrated by the two narrow LDOS peaks
around the Fermi level ($E_F$). At finite $\nu$, the bands around $E_F$ are still rather flat, while the LDOS peaks are wider
than those of the single-particle LDOS.
Moreover, three LDOS peaks can be seen. For positive $\nu$, the peak value above $E_F$ becomes smaller with more filled bands, while
the peak just below $E_F$ becomes higher. The LDOS peaks for negative $\nu$ show a similar trend.
Such evolution behavior of the LDOS peaks with $\nu$ is roughly consistent with the general trend of the
scanning tunnelling spectrum observations on TBG with varying doping levels\cite{xie2019spectroscopic}.
In particular, three spectra peaks are observed in experiments away from CNP, while the peak separations
are smaller than those for the calculated LDOS at the MF level.

\begin{figure}[t]
\includegraphics[width=1\columnwidth]{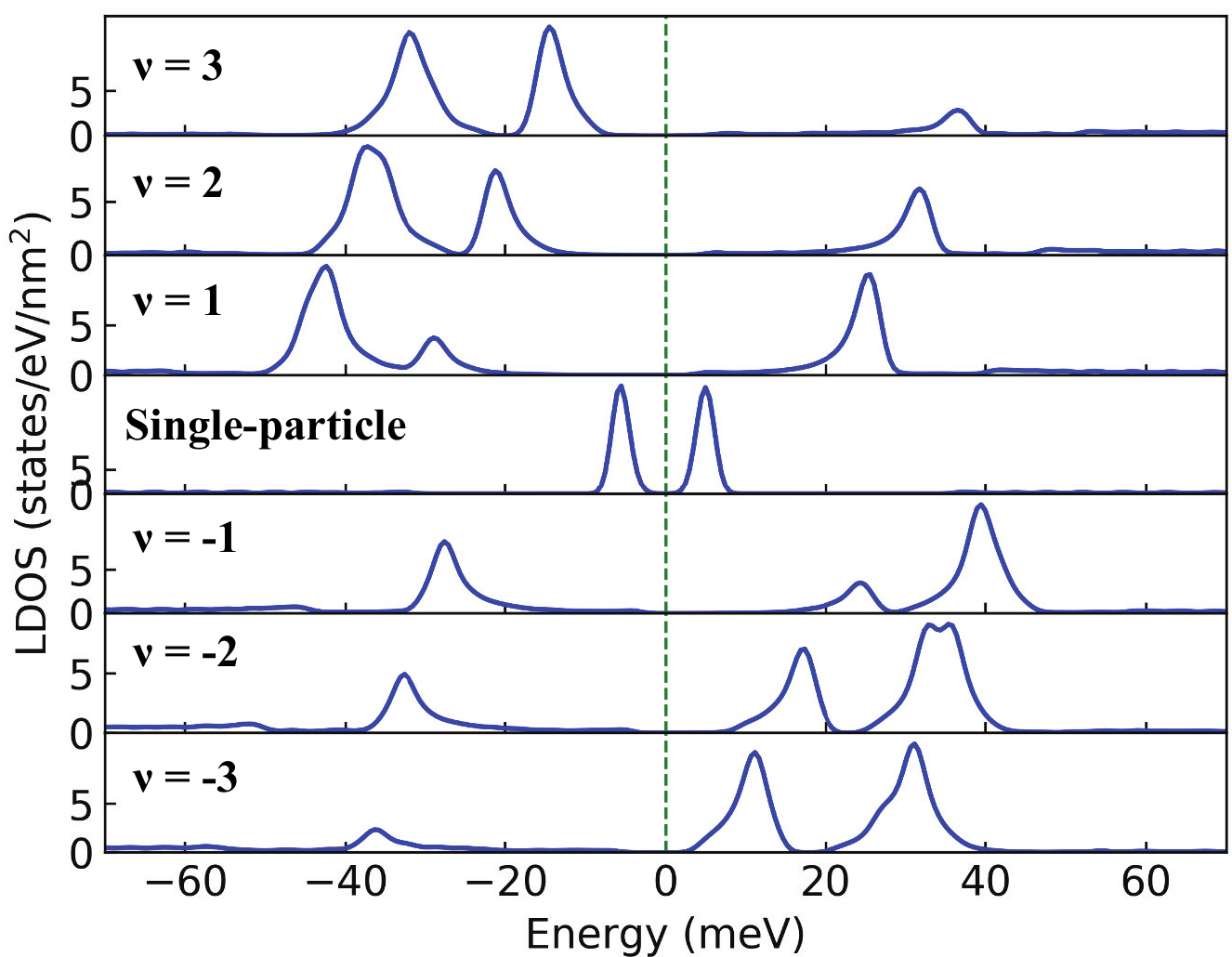}
\caption{(Color online) The local density of states (LDOS) at the AA-stacked part of
G3/G2 for the MF band states at each finite $\nu$ and the single-particle band states at CNP.
The Fermi levels
are set to be zero and are represented by the dashed green lines.
\label{fig6}}
\end{figure}

\section{Summary and Conclusions}

Full relaxation of the commensurate double moir\'{e} superlattices in TBG/BN has been performed.
The superlattice in G2/BN with a twist angle of about 1.6$^\circ$ becomes commensurate with that in G3/G2 with the magic angle $\theta_m$.
The inversion-asymmetric spatial variation of the interlayer interaction potential between G2 and BN leads to
the broken $C_2$ symmetry in the structural deformation in G2, while the symmetry is approximately maintained in G3.
For the relaxed TBG/BN, an effective single-particle Hamiltonian $\hat{H}^0$ taking into account the relaxation effect
and the full effective Hamiltonian introduced by BN
is built for the
moir\'{e} superlattice in G3/G2.
The Hamiltonian terms in $\hat{H}^0$ induced by both relaxation and BN break the $C_2$ symmetry,
leading to an opened gap at CNP that separates the flat conduction and valence bands.
These gapped flat bands have finite Chern numbers.
The symmetry breaking in $\hat{H}^0$ is found to mainly originate from the intralayer inversion-asymmetric
strain fields in G2, which induce spatially non-uniform modifications of the intralayer on-site energies and the
nearest-neighbor hopping terms.
The influence on the symmetry of the interlayer hopping by the structural deformation in G2 is much weaker than
that for the intralayer terms.
The MF band structures of the SCHF ground states at different filling factor $\nu$ are acquired based on $\hat{H}^0$ in the plane-wave-like basis.
For TBG/BN with $\theta_m$, the ground states with $|\nu|$ = 1-3 are all insulating with narrow MF gaps.
When the flat conduction bands are filled, the gap at $\nu$ = 1 ($\nu$ = -1) is smaller than that at $\nu$ = 3 ($\nu$ = -3).
Our study thus suggests that TBG/BN is a promising platform for observation of
the nontrivial topological properties driven by the broken symmetry.

\label{Acknowledgments}
\begin{acknowledgments}
We gratefully acknowledge valuable discussions with D. Tom\'anek, D. Liu,
H. Xiong, and Q. Zhang.
This research was supported by
the National Natural Science Foundation of China (Grants No. 11974312 and No. 11774195),
and the National Key Research and Development Program of China(Grant No. 2016YFB0700102).
The calculations were performed on TianHe-1(A) at National Supercomputer Center in Tianjin.
\end{acknowledgments}

\section*{Appendix}
\setcounter{equation}{0}
\renewcommand{\theequation}{A\arabic{equation}}

\subsection{Geometry of the double moir\'{e} superlattices in TBG/BN}

\begin{figure}[tb]
\includegraphics[width=1.0\columnwidth]{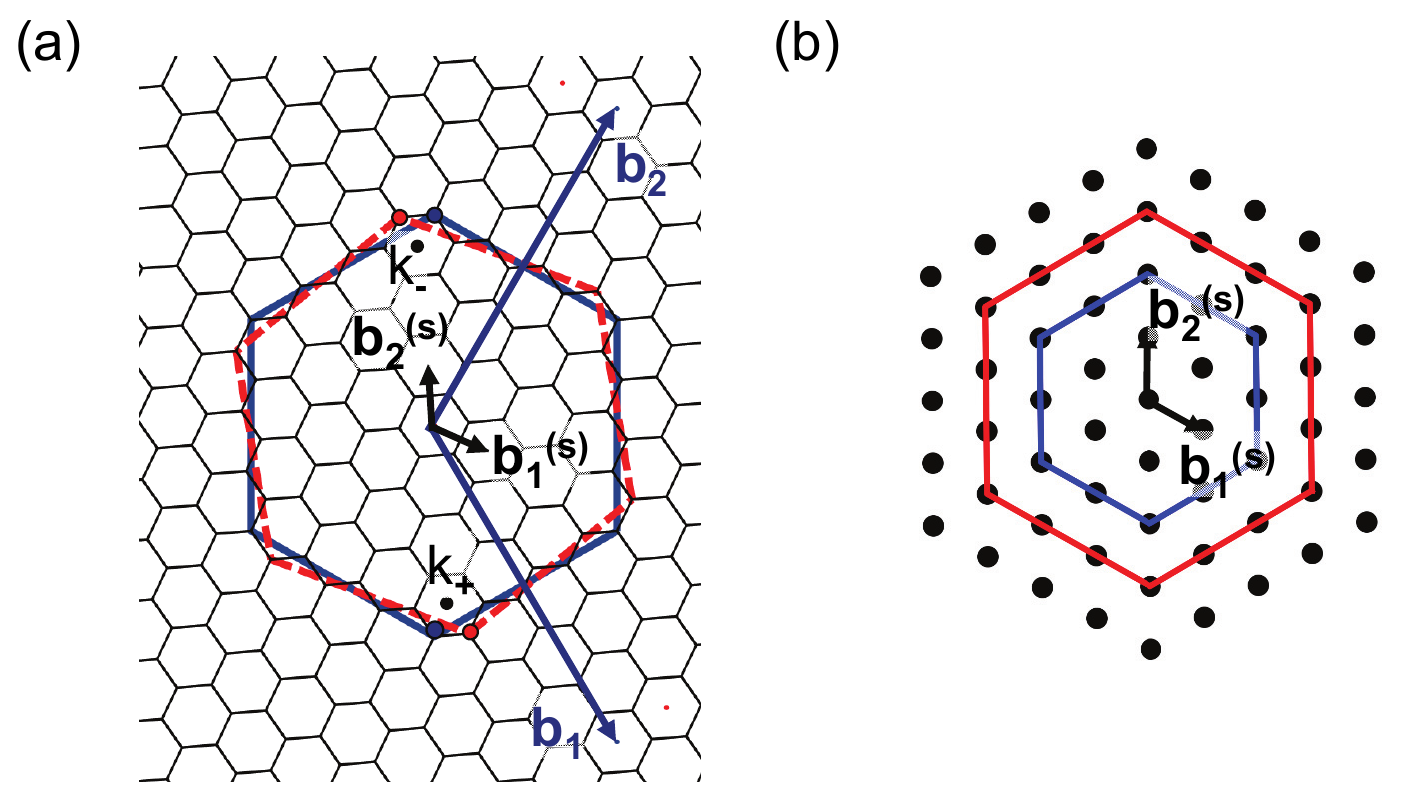}
\caption{(Color online).
(a) The schematic reciprocal lattice of the moir\'{e} superlattice in G3/G2 of TBG/BN.
Small hexagons are periodic BZs of the superlattice, spanned by
${\bf{b_1^{(s)}}}$ and ${\bf{b_2^{(s)}}}$.
Large hexagons are BZ of the fixed layer (solid), spanned by $\bf{b}_1$
and $\bf{b}_2$, and BZ of the twisted layer (dashed).
$\mb{k}_\xi$ ($\xi=\pm$) is the center
of one of the supercell BZs containing the Dirac points of the fixed and twisted layers at their
corners in the valley $\xi$, and we take $\mb{k}_{-} = - \mb{k}_+$.
(b) The reciprocal lattice vectors of the superlattice.
Inside the red hexagon are the 37 vectors with magnitudes smaller than or equal to $3|{\bf{b_1^{(s)}}}|$,
and the 19 vectors inside the blue hexagon have magnitudes smaller than or equal to $2|{\bf{b_1^{(s)}}}|$.
\label{fig7}}
\end{figure}

We consider the trilayer structures with the top TBG nearly aligned with
the bottom BN layer. The top graphene layer (G3) and the bottom BN layer are rotated by
$\theta$ and $\theta'$ counterclockwise respectively with respect to the fixed middle graphene layer (G2),
as shown schematically in Fig. 1(a). Due to the relative twist between the adjacent layers and the lattice-constant
mismatch between graphene and BN, double moir\'{e} superlattices are formed in TBG/BN, as shown in Fig. 1(b).
As directly seen from the moir\'{e} patters with small twist angles,
the spanning vectors ($\mb{a_j^{(s)}}$ with $j = 1, 2$) of the moir\'{e} superlattice in G3 on G2 (G3/G2)
are along the armchair directions of graphene, while those ($\mb{a_j^{(s)'}}$) of
the moir\'{e} superlattice in G2 on BN (G2/BN) are along the zigzag directions.
Here, we study the commensurate double moir\'{e} superlattices with small twist angles, then
$a^{(s)} \simeq \sqrt{3} a^{(s)'}$ for the minimum commensurate supercells, as shown in Fig. 1(b), where $a^{(s)}$ and $ a^{(s)'}$
are lengthes of $\mb{a_j^{(s)}}$ and $\mb{a_j^{(s)'}}$, respectively.
In the following, the geometry parameters of the commensurate superlattices are obtained.

The unit cell of the fixed G2 layer is spanned by the basis vectors $\mb{a_1} = a(\sqrt{3}/2, -1/2)^\mathrm{T}$ and
$\mb{a_2} = a(\sqrt{3}/2, 1/2)^\mathrm{T}$, where the superscript T denotes matrix transposition. The basis vectors of the BN layer
become
$\mb{a'_j} = S \mb{a_j}$ ($j$ = 1, 2), where the transformation matrix
\begin{equation}
S = \frac{1}{1 + \epsilon} \left( {\begin{array}{*{20}{c}}
 \cos\theta'  & -\sin\theta'   \\
 \sin\theta' & \cos\theta'
\end{array}} \right).
\end{equation}
In the G3 layer, the unit cell is spanned by $T_{\theta} \mb{a_j}$ ($j$ = 1, 2),
where $T_{\theta}$ denotes the counterclockwise rotation by $\theta$.

In G3/G2, the strictly periodic moir\'{e} superlattices are spanned by the basis vectors
$\mb{a_1^{(s)}} = N \mb{a_1} + (N+1) \mb{a_2}$ and $\mb{a_2^{(s)}} = -(N+1) \mb{a_1} + (2N+1) \mb{a_2}$, where $N$ is an integer.
The relation between $\theta$ and $N$ can be expressed as
$\cos \theta = (1+6N+6N^2)/(2+6N+6N^2)$. For $N = 30$, $\theta$
takes 1.0845$^\circ$ and is closest to the experimentally observed $\theta_m$.
The length of $\mb{a_j^{(s)}}$ is given by $a^{(s)} = a/[2\sin(\theta/2)]$.
In G2/BN, the spanning vectors of the moir\'{e} superlattices
can be taken as $(S^{-1} - I)\mb{a_j^{(s)}} = \mb{a_j}$, which gives the
moir\'{e} period $a^{(s)'} = a/\sqrt{\epsilon^2 + 4(1+\epsilon)\sin^2(\theta'/2)}$.
Since $a^{(s)} \simeq \sqrt{3} a^{(s)'}$ in the commensurate superlattices,
$\theta' \simeq 1.6^\circ$ for $\theta$ around $\theta_m$ and
$\epsilon = -1.70\%$.
With the approximate $\theta'$, the basis vectors of BN $\mb{a'_j}$ can be computed.
For $N$ around $30$, we find that $\mb{a_1^{(s)}}$ can be approximately expressed as
$(N+1) \mb{a'_1} + (N-1) \mb{a'_2}$, and it can become an exact lattice vector of
BN by varying $\theta'$ around $1.6^\circ$ and varying $\epsilon$ around $-1.70\%$ very slightly.
Then in the strictly periodic  moir\'{e} superlattices,
$\mb{a_1^{(s)}} = N \mb{a_1} + (N+1) \mb{a_2} = (N+1) \mb{a'_1} + (N-1) \mb{a'_2}$,
which gives
\begin{eqnarray}
\epsilon &=& \sqrt{\frac{1+3N^2}{1+3N+3N^2}} - 1, \\
  \tan \theta' &=& \frac{\sqrt{3}(3N+1)}{6N^2+3N-1}.
\end{eqnarray}
The computed $\epsilon$ and $\theta'$ for the three considered cases with $N$ from 28 to 30 are listed in
Table I. We note that for $N$ outside this range, the deviation of $\epsilon$ from
$-1.70\%$ becomes rather large.
In addition, when BN is rotated clockwise with respect to G2, the
deviation of $\epsilon$ from
$-1.70\%$ for the commensurate superlattices is also rather large.

The sublattice-A and sublattice-B atoms in a unit cell of G2 are located at
$(\mb{a_1} + \mb{a_2})/3$ and $(2\mb{a_1} + 2\mb{a_2})/3$, respectively.
In BN, the lattices formed by the boron and nitrogen atoms are labeled as sublattice-A and
sublattice-B, respectively.
For the initial structure without twist and lattice-constant mismatch, the adjacent layers are all taken to be
AA-stacked.
In TBG/BN, the local stackings between adjacent layers
vary continuously and are characterized by the relative shift vectors.
At an in-plane position $\mb{r}$ in the rigid superlattice, the shift vector between G3 and G2 is taken to be
$\bm{\delta} = (I - T_{-\theta}) \mb{r}$, and
the shift vector between G2 and BN is given by $\bm{\delta'} = (S^{-1} - I) \mb{r}$.

\subsection{Relaxation of TBG/BN by Euler-Lagrange equations}

The in-plane structural relaxation in each layer of TBG/BN is driven by the
interlayer interaction potentials, which vary with the local stacking configurations.
At an in-plane position in the superlattice, the local spacings between the adjacent layers are
taken to be the optimal values for the local stacking configurations, and the local potentials are taken as
the calculated energies of the corresponding bilayers.
Between G3 and G2, the interlayer potential $V$ as a function of the local shift vector $\bm{\delta}$
can be expressed as $V(\bm{\delta}) = \tilde{V} \sum_{j=1}^{3} \cos(\mb{G}_{j} \cdot \bm{\delta})$, where the sum is
limited to three shortest vectors, $\mb{G_1} = \mb{b_1}$,
$\mb{G_2} = \mb{b_2}$, and $\mb{G_3} = -\mb{b_1}-\mb{b_2}$,
and the parameter obtained from \emph{ab-initio} calculations is
$\tilde{V} = 0.817$ meV/{\AA}$^2$.
The $ab-initio$ computational approach is detailed in Appendix C.
It is noted that $V(\bm{\delta})$ has the inversion symmetry with
$V(\bm{\delta}) = V(-\bm{\delta})$.
However, the inversion symmetry is absent in the interlayer potential $V'$
between G2 and BN as the two sublattices in BN are composed of different kinds of
atoms. The $V'$ is thus expressed as
$V'(\bm{\delta}) = \tilde{V}' \sum_{j=1}^{3} \cos(\mb{G}_{j} \cdot \bm{\delta} + \phi_V')$,
where $\tilde{V}' = 0.845$ meV/{\AA}$^2$ and $\phi_V' = -50.26^\circ$.

We have employed the continuum elastic theory to evaluate $E_{tot}$ of a supercell.
$E_{tot}$ as the sum of the elastic energy ($E_{el}$) in each layer and the
interlayer interaction energy ($E_{int}$) is a functional of the displacement fields
$\mb{u^{(n)}}(\mb{r})$ with $n$ the layer index shown in Fig. 1(a).
The elastic energy functional is given by\cite{Andres2012}
\begin{eqnarray}
E_{el} = \! \sum_{n=1}^{3} \int d{\bf{r}} \biggl\{ \frac{\lambda_n+\mu_n}{2}
            \left(\frac{\partial u_x^{(n)}}{\partial x} \!+\!
            \frac{\partial u_y^{(n)}}{\partial y}\right)^2
            + \nonumber \\
\!\frac{\mu_n}{2} \left[
         \left(\frac{\partial u_x^{(n)}}{\partial x} -
          \frac{\partial u_y^{(n)}}{\partial y}\right)^2 %
          \!\!\! + \!
         \left(\frac{\partial u_y^{(n)}}{\partial x} +
          \frac{\partial u_x^{(n)}}{\partial y}\right)^2
          \right] \biggr\} ,%
\label{eq1}
\end{eqnarray}
where the integral extends over a moir\'{e} supercell.
The calculated 2D elastic Lam\'{e} factors are
$\lambda_1=1.779$~eV/{\AA}$^2$ and $\mu_1=7.939$~eV/{\AA}$^2$ for BN and
$\lambda_2=\lambda_3=3.653$~eV/{\AA}$^2$ and $\mu_2=\mu_3=9.125$~eV/{\AA}$^2$ for graphene.
The $E_{int}$ is given by the integral of the local interlayer interaction potentials\cite{Zhou2015}
\begin{equation}
E_{int} = \int \{V[\bm{\delta}(\mb{r})] +  V'[\bm{\delta}'(\mb{r})] \}d{\mb{r}},
\end{equation}
where $\bm{\delta}(\mb{r}) = (I - T_{-\theta}) \mb{r}  + \mb{u}^{(3)}(\mb{r}) - \mb{u}^{(2)}(\mb{r})$ and
$\bm{\delta}'(\mb{r}) = (S^{-1} - I)\mb{r}  + \mb{u}^{(2)}(\mb{r}) - \mb{u}^{(1)}(\mb{r})$
for the relaxed structure.

The $E_{tot}$ of a supercell
can be expressed as
$E_{tot} = \int L[\mb{u}^{(1)}, \mb{u}^{(2)}, \mb{u}^{(3)}] d{\mb{r}}$.
The minimization of $E_{tot}$ as a functional of $\mb{u}^{(n)}$ leads to
a series of Euler-Lagrange equations\cite{Nam2017}
\begin{equation}
\frac{\partial}{\partial x}\left[\frac{\partial L}{\partial (\partial u^{(n)}_\nu/\partial x)}\right] +
\frac{\partial}{\partial y}\left[\frac{\partial L}{\partial (\partial u^{(n)}_\nu/\partial y)}\right] - \frac{\partial L}{\partial u^{(n)}_\nu} = 0
\end{equation}
where $\nu = x, y$. To solve these equations,
$\bf{u}^{(n)}(\mb{r})$ is expanded in Fourier series as
\begin{equation}
\bf{u}^{(n)}({\bf{r}}) = %
\sum_{\bf{G}^{(s)}} \bf{\tilde{u}}^{(n)} ({\bf{G}}^{(s)})
e^{i{\bf{G}}^{(s)}{\cdot}{\bf{r}}},
\end{equation}
where the summation is over nonzero reciprocal lattice vectors $\bf{G}^{(s)}$ of the supercell.
The $\partial V/\partial \bm{\delta}$ and $\partial V'/\partial \bm{\delta}'$ are also expanded as
$\sum_{\mb{G^{(s)}}} \mb{\tilde{f}}(\mb{G^{(s)}}) e^{i \mb{G^{(s)}} \cdot \mb{r}}$
and $\sum_{\mb{G^{(s)}}} \mb{\tilde{f}}'(\mb{G^{(s)}}) e^{i \mb{G^{(s)}} \cdot \mb{r}}$, respectively.
Substitution of these
Fourier expansions into Eq. (A6) leads to
\begin{eqnarray}
-\left(\begin{matrix}
   (\lambda_n+2\mu_n)q^2_x + \mu_n q^2_y  & (\lambda_n+\mu_n)q_x q_y \cr
   (\lambda_n+\mu_n)q_x q_y & (\lambda_n+2\mu_n)q^2_y + \mu_n q^2_x \cr
   \end{matrix} \right)&
   \left(\begin{matrix}
   u^{(n)}_x(\mb{q}) \cr
   u^{(n)}_y(\mb{q}) \cr
   \end{matrix} \right) \nonumber \\
= \tilde{F}_n({\mb{q}}),\ \ \
\end{eqnarray}
where $\mb{q}$ takes each $\mb{G^{(s)}}$,
$\tilde{F}_1 = (-\tilde{f}'_x, -\tilde{f}'_y)^\mathrm{T}$,
$\tilde{F}_2 = (\tilde{f}'_x - \tilde{f}_x, \tilde{f}'_y - \tilde{f}_y)^\mathrm{T}$,
and $\tilde{F}_3 = (\tilde{f}_x, \tilde{f}_y)^\mathrm{T}$.

\subsection{\emph{Ab-initio} computational approach}

The \emph{ab-initio} density functional theory (DFT) calculations of the shifted bilayers
are performed
using the VASP code\cite{Efficiency1996Jul,Efficient1996Oct} to extract the required model parameters.
The local density approximation (LDA)\cite{SELF-INTERACTION1981}
functional is adopted.
No van der Waals (vdW) functional has been used, while the vdW interaction is partly accounted by the LDA functional\cite{Jung2014}.
The projector augmented wave (PAW) potentials\cite{PROJECTOR1994Dec,From1999Jan}
are used with a kinetic energy cutoff of 600 eV.
The BZ sampling is done using
a 36 $\times$ 36 $\times$ 1 Monkhorst-Pack (MP) grid\cite{SPECIAL1976}.
The vacuums in the $z$ direction are larger than 17 {\AA}.
The tolerance for the energy convergence is 10$^{-6}$ eV.
The optimal interlayer distances ($h$) of the shifted bilayers are obtained by minimizing the total energies through
adaptively scanning $h$ with a final precision of 0.001 {\AA}.


%

\end{document}